\documentclass[letter]{aa}

\usepackage{txfonts}

\usepackage{graphicx}	
\usepackage{amsmath}	
\usepackage{amssymb}	
\usepackage{gensymb}
\usepackage{tabulary}
\usepackage{subcaption}
\usepackage{epstopdf}
\usepackage{natbib}
\usepackage{afterpage}
\usepackage{float}
\usepackage{booktabs}
\usepackage[space]{grffile}
\usepackage[percent]{overpic}
\usepackage{xcolor, soul}
\usepackage{hyperref}
\sethlcolor{white}

\begin{document}


\title{Complete census of massive slow rotators in ten large galaxy clusters}

\author{Mark T. Graham$^1$\thanks{E-mail: mark.graham@physics.ox.ac.uk}
\and Michele Cappellari$^1$
\and Matthew A. Bershady$^{2,3}$
\and Niv Drory$^4$}
\authorrunning{M. T. Graham et al.}
\institute{Sub-department of Astrophysics, Department of Physics, University of Oxford, Denys Wilkinson Building, Keble Road, Oxford, OX1 3RH, UK
\and Department of Astronomy, University of Wisconsin-Madison, 475N. Charter St., Madison WI 53703, USA
\and South African Astronomical Observatory, Cape Town, South Africa
\and McDonald Observatory, The University of Texas at Austin, 1 University Station, Austin, TX 78712, USA
}
\abstract{
Galaxy interactions leave imprints in the motions of their stars, and so observing the two-dimensional stellar kinematics allows us to uncover their formation process. Slow rotators, which have stellar orbits dominated by random motions, are thought to be the fossil relics of a sequence of multiple gas-poor mergers, in an environment where the cold gas required to form new stars is nearly absent. Indeed, observations of a handful of nearby galaxy clusters have indicated that slow rotators are preferentially found in the gas-poor, dense cores of clusters, which themselves must form by merging of smaller groups. However, the generality of this result and connection between kinematics and environment is currently unclear, as recent studies have suggested that, at given stellar mass, the environment does not influence the formation of slow rotators. Here we address this issue by combining a careful quality-assessed sample selection with two-dimensional stellar kinematics from a large galaxy survey and a novel photometric classification approach where kinematics are unavailable. We obtain the first complete census of the location of massive slow rotators in ten large clusters: in all cases, slow rotators are extremely rare and generally trace the clusters density peaks. This result unambiguously establishes that massive slow rotators are the relics of violent hierarchical cluster formation.}


\maketitle

We conduct the first complete census of kinematic morphology for a subsample of 2007 galaxies taken from the catalogue presented in \cite{graham2019bcatalogue} of about 14,000 galaxies (approximate minimum stellar mass $M \gtrsim 7.4 \times 10^{9} \textrm{M}_{\odot}$, redshift $z \lesssim 0.08$). The galaxies in this subsample of 2007 galaxies are assigned to a cluster using the robust local neighbour finder \texttt{TD-ENCLOSER}, described in \cite{graham2019atechnical}. All of the galaxies have photometric properties such as apparent magnitude, and 1636 ($\sim82\%$) have spectroscopic redshifts. 193 galaxies ($\sim10\%$) in the sample have stellar kinematics obtained by the SDSS \citep{gunn20062, blanton2017sloan, smee2013multi} MaNGA survey \citep{bundy2015overview, drory2015manga, law2015observing, law2016data, yan2016spectrophotometric, yan2016sdss, wake2017sdss}, and a further 25 galaxies in the Coma cluster have integral-field kinematics obtained by with other instruments (23 from SWIFT; \citealp{thatte2006oxford, houghton2013densest} and two from the MASSIVE survey; \citealp{veale2016massive, ene2018misalignment}). Galaxies can be robustly separated into fast and slow rotators (FR and SR) using a proxy for specific stellar angular momentum $\lambda_{R_e}$, derived from the stellar kinematics, the apparent flattening $\epsilon$ and the stellar mass. Of the remaining 1791 galaxies for which we do not have stellar kinematics, only 123 ($\sim7\%$) satisfy the criteria for SRs: $\rm{M}_{\rm{\ast}}\geq \rm{M}_{\rm{crit}} \equiv 2\times10^{11}\rm{ M}_{\odot}$ and $\epsilon<0.4$. We classify these 123 galaxies as SR/FR \textit{candidates} according to their visual morphology. We showed that one can guess an early type galaxy (ETG) that is more massive than $\rm{M}_{\rm{crit}}$ to be a FR with $\sim92\%$ accuracy, and so of the 1964 \textit{total} FRs in the sample, we expect just \textit{nine} ($\sim0.5\%$) to be misclassified, as all galaxies less massive than $\rm{M}_{\rm{crit}}$ are by definition fast rotators \citep{cappellari2016structure}. Hence, we can be certain that almost \textit{all} galaxies classified as FRs in our sample are \textit{genuine} FRs. On the other hand, given a guess of SR, we are correct in just under half of all cases ($\sim48\%$), meaning that the lower and upper limit for the number of massive SRs per cluster is given by the number of confirmed and confirmed + candidate SRs respectively.\par
In Figures 1 and 2, we present the on-sky projections for our cluster sample, clearly highlighting the location of the massive SRs. The key result is that the central part of \textit{every} cluster has \textit{at least} one massive SR, and almost all massive SRs that are not in the centres are associated with substructure (e.g. Abell 2065). Abell 2197 is known to have a bimodal galaxy distribution \citep{rines2002abell}, and Abell 1775 is known to have multiple components \citep{oegerle1995clusters, kopylov2009a1775, zhang2011a1775}; in these cases, a massive SR exists at each density peak. In cases where clusters appear elongated and are relaxing post merger, the SRs are aligned with the orientation of the cluster major axis, suggesting they sit at the bottom of the potential well in their respective subclusters (e.g. Abell 2167, \citealp{marini2004corona}; Abell 2161, \citealp{rines2006cluster}; Abell 2079 and Abell 1314, \citealp{wilbur2018clusters}).\par
Our results are in good agreement with early assessments for three clusters \citep{cappellari2016structure} using data compiled from observations of the Virgo \citep{cappellari2011atlas3d}, Fornax \citep{scott2014distribution} and Coma \citep{houghton2013densest} clusters. Our results are consistent with the theoretical viewpoint that the progenitors of present-day core SRs formed at the centres of the most massive haloes early in the history of the Universe, and experienced successive dry (gas-poor) mergers as a result of merging with other galaxy groups. Our findings are also consistent with results from eight clusters in the Southern sky \citep{brough2017kinematic} which show that in general, the most massive ETGs with low angular momentum are found in the dense cluster core. However, that study was based on very incomplete coverage of the clusters. Moreover, the same study, along with two other recent studies \citep{veale2017angular, greene2017kinematic} have shown that at \textit{fixed stellar mass}, there is little or no trend in the observed angular momentum or the fraction of SRs with environment, which is a conclusion not supported by this work.\par
If present-day massive SRs did form by hierarchical merging, then we should indeed observe the stellar mass to be the main driver of slow rotation in galaxies since stellar mass cannot disappear and can be easily be determined observationally. We would still expect to see a trend with environment as the frequency of minor mergers must increase with increasing galaxy density. However, we would expect this trend to be weaker than the one with stellar mass because the environment is non-trivial and challenging to measure reliably and the boundaries of clusters and groups are not easy to define. Moreover, massive ellipticals ``slosh'' about within clusters \citep{barbosa2018sloshing}, especially during or after mergers of group haloes \citep{randall2010cluster}, and so may appear in lower density environments in a later stage as a result. Finally, it is possible that a present-day SR may have accreted a large proportion of its neighbours and so may appear to be in a low-density environment or even in isolation. These effects conspire to wash out the dependence of angular momentum on environment for SRs at low redshift which constitute the evolutionary end points. However, if taken at face value, the lack of such a trend implies that the environment \textit{itself} has little to no effect on the formation of SRs.\par
One crucial aspect of massive SRs is that they are intrinsically rare, making up between 21-35 ($\sim1\%-2\%$) of the galaxies in our cluster sample. Thus, they are vastly outnumbered by FRs and spirals and so any trends for subsamples containing galaxies less massive than $M_{\rm{crit}}$ will be driven by these galaxies. Indeed, FRs are necessary in order to build up the number density in the cores of clusters and so on average, only one in four galaxies will be a SR at a given number density. These aspects should be taken into account when interpreting trends at fixed stellar mass. Furthermore, we have used a larger sample with groups of 10 or more members to show that massive SRs are biased towards the densest regions of groups and clusters \textit{particular to each cluster}. In other words, the fraction of SRs is \textit{not} an absolute function of number density, but \textit{is} a function of the \textit{relative} number density compared to the maximum in each cluster. This means that by combining multiple clusters in a single kinematic morphology-density relation, the profiles of increasing F(SR) with number density, which represent what we see by eye, are smoothed out.\par
Our study has provided the first complete view of the relationship between kinematic morphology, which encodes the fossil record of a galaxy's formation mechanism, and environment, which strongly influences a galaxy's evolution. We have shown that within the cluster environment, massive SRs are almost non-existent beyond about 0.5 Mpc from the cluster centre or local overdensity. Hierarchical formation predicts that the most massive clusters, such as the ones presented here, are built up from smaller groups. The fact that the distribution of massive SRs in present-day clusters is concentrated towards the density peaks provides conclusive evidence that the progenitor haloes (``building blocks'') each contained massive SRs or passive FRs at their centres. The results of this study provide a unique benchmark for future theoretical studies of galaxy formation and present a challenge to cosmological simulations which must be able to reproduce what we see by eye.
\begin{figure*}
  \centering
\vspace{-0.13em}

    \begin{subfigure}[b]{0.48\linewidth}
      \begin{overpic}[width=\linewidth]{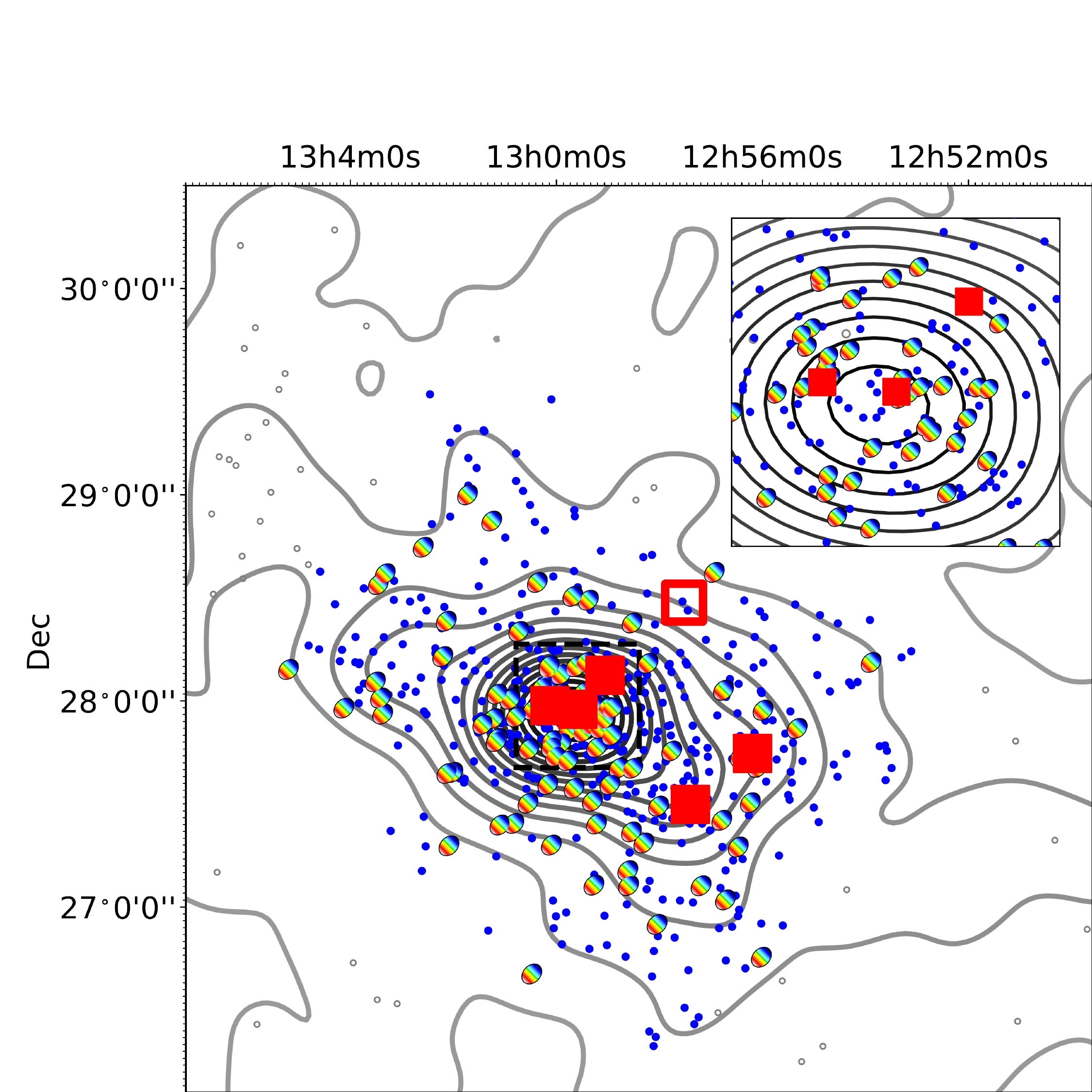}
        \put (20,76) {\fontsize{8}{0}\hl{Coma Cluster (Abell 1656)}}        \put (20,71) {\fontsize{8}{0}\hl{$\mathcal{N}=509$, $z\approx0.023$}}
      \end{overpic}
    \end{subfigure}
    \hspace{-0.7em}
    \begin{subfigure}[b]{0.48\linewidth}
      \begin{overpic}[width=\linewidth]{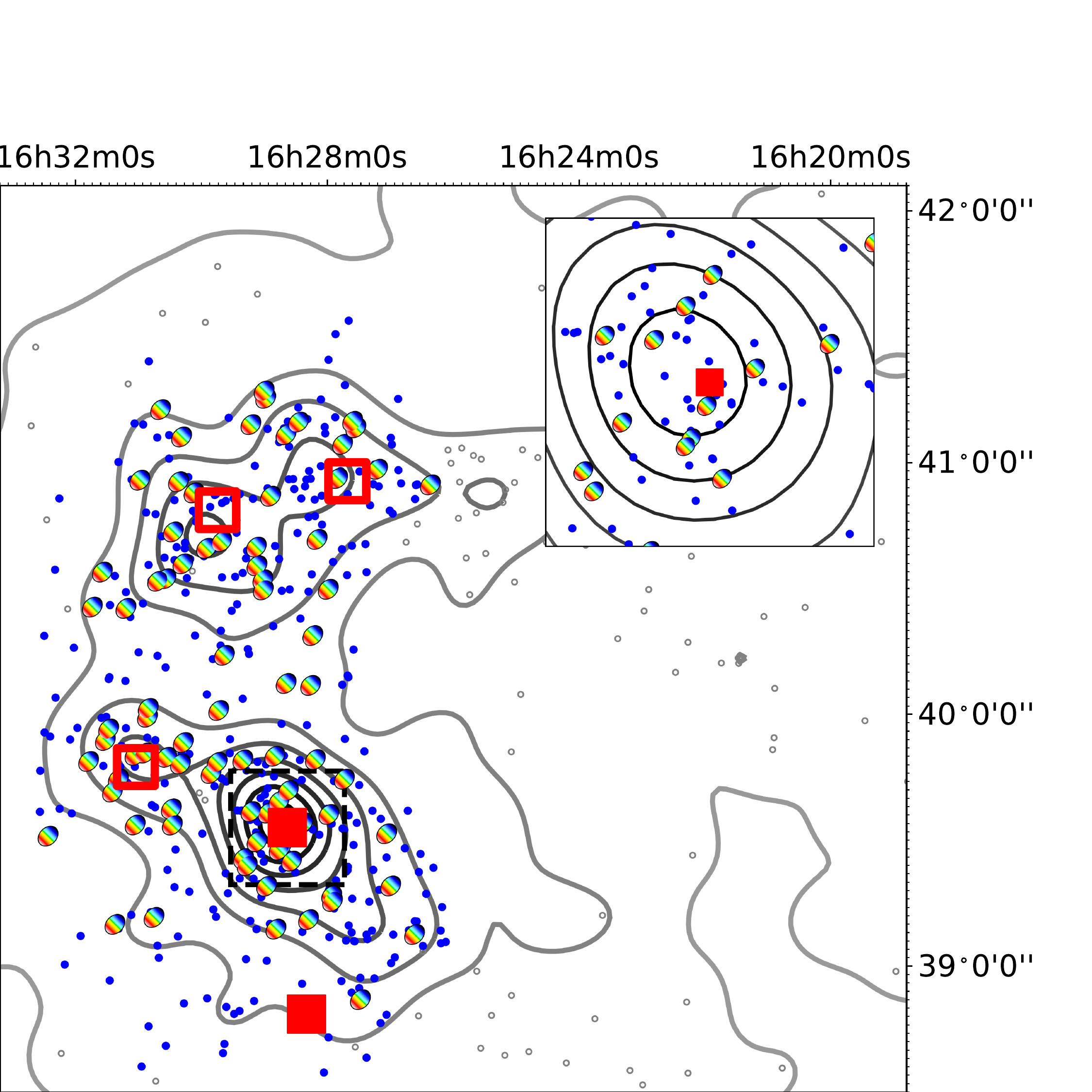}
        \put (3,76) {\fontsize{8}{0}\hl{A2199 ($\downarrow$) and A2197 ($\uparrow$)}}        \put (3,71) {\fontsize{8}{0}\hl{$\mathcal{N}=425$, $z\approx0.03$}}
      \end{overpic}
    \end{subfigure}
    \hspace{-0.7em}
    \begin{subfigure}[b]{0.48\linewidth}
      \begin{overpic}[width=\linewidth]{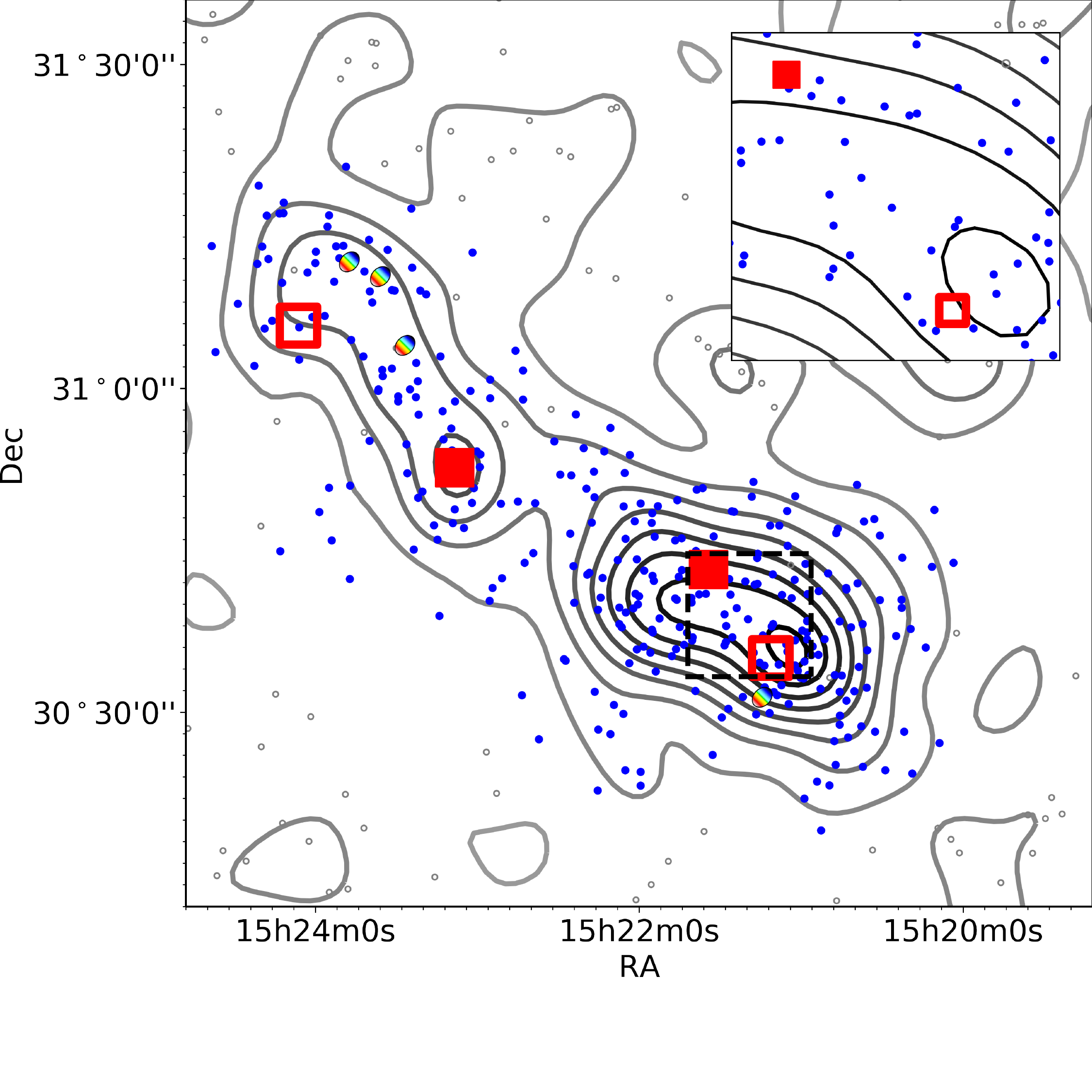}
        \put (20,93) {\fontsize{8}{0}\hl{A2061 ($\rightarrow$) and A2067 ($\leftarrow$)}}        \put (20,88) {\fontsize{8}{0}\hl{$\mathcal{N}=320$, $z\approx0.07$}}
      \end{overpic}
    \end{subfigure}
    \hspace{-0.7em}
    \begin{subfigure}[b]{0.48\linewidth}
      \begin{overpic}[width=\linewidth]{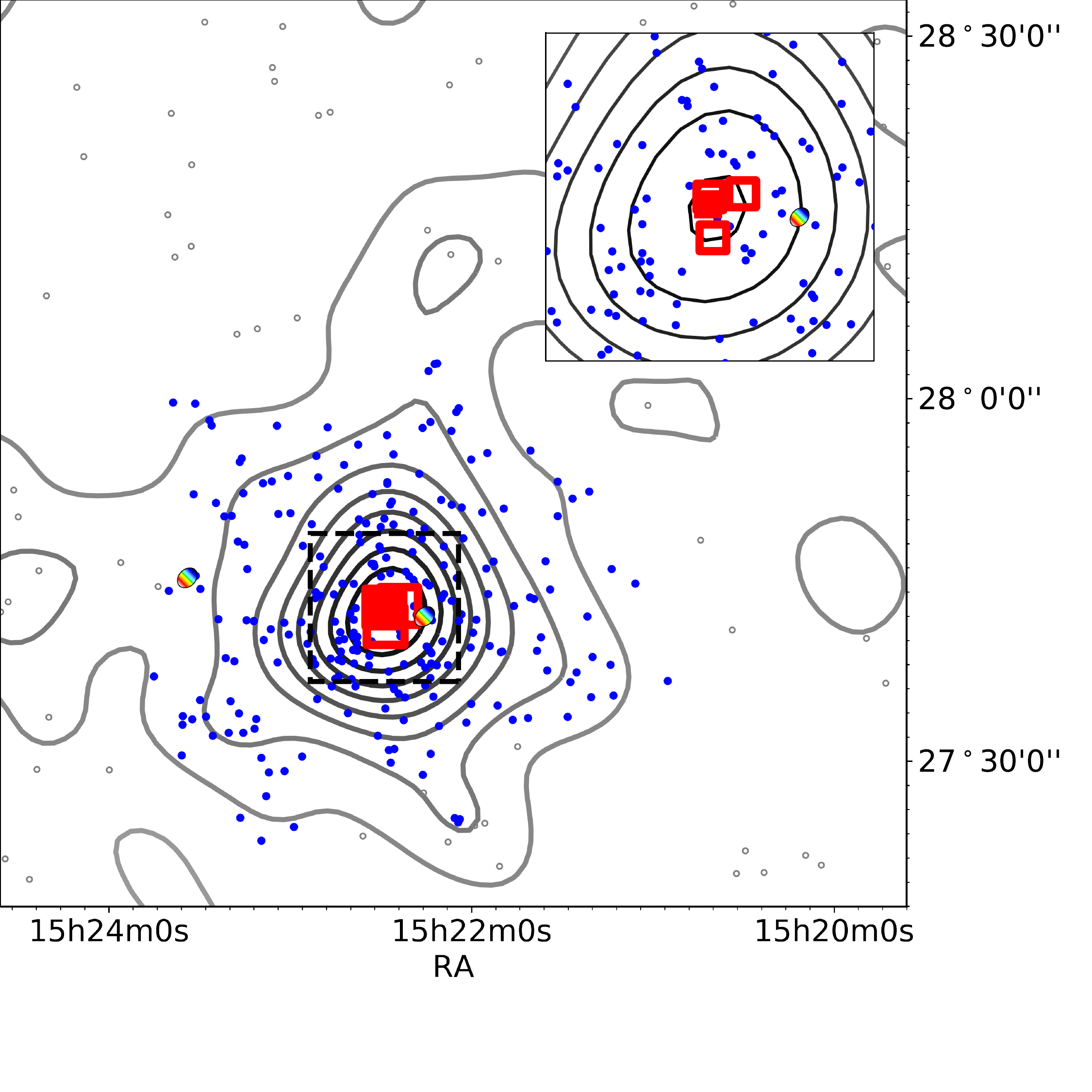}
        \put (3,93) {\fontsize{8}{0}\hl{A2065}}        \put (3,88) {\fontsize{8}{0}\hl{$\mathcal{N}=254$, $z\approx0.07$}}
      \end{overpic}
    \end{subfigure}
    \hspace{-0.7em}
    \vspace{-1em}
\caption[Galaxy clusters where at least one galaxy has been observed with kinematics: Coma cluster (A1656), A2197, A2199, A2061, A2067, A2065]{\textbf{Galaxy clusters where at least one galaxy has been observed with kinematics}. Cluster members are indicated by either a red square if the galaxy is a SR with $M>M_{crit}$, or a blue circle if otherwise. Filled points are galaxies which are confirmed as FRs or SRs with kinematics, whereas unfilled points are galaxies which MG and MC have classified visually. The contour levels indicate increasing density with increasing opacity. Galaxies which are not cluster members are shown as grey unfilled circles. The square has an area equal to 1 Mpc$^2$ and is centred on the densest peak in the field of view. A close up of the region within the square is shown in the inset. For each cluster, we give the designation, number of members $\mathcal{N}$ and approximate redshift. \textit{Top Left}: The Coma Cluster (A1656). We supplement our classifications with published classifications \citep{houghton2013densest,veale2016massive, ene2018misalignment}, including the three SRs at the core. \textit{Top Right}: A2197 (top) and A2199 (bottom). Abell 2197 has a bimodal velocity distribution and is considered to be two clusters, A2197W and A2197E, which may be interacting \citep{rines2002abell}. \textit{Bottom Left}: A2061 (lower right) and A2067 (upper left). This pair of clusters is in the Corona Borealis Supercluster \citep{postman1988corona, marini2004corona, pearson2014corona} (CBS) and is likely to be bound as the velocity difference is only $\sim$600 km s$^{-1}$ \citep{rines2006cluster}. \textit{Bottom Right}: A2065. Part of the CBS.}
\end{figure*}
\begin{figure*}
  \centering
\vspace{-0.13em}

    \begin{subfigure}[b]{0.48\linewidth}
      \begin{overpic}[width=\linewidth]{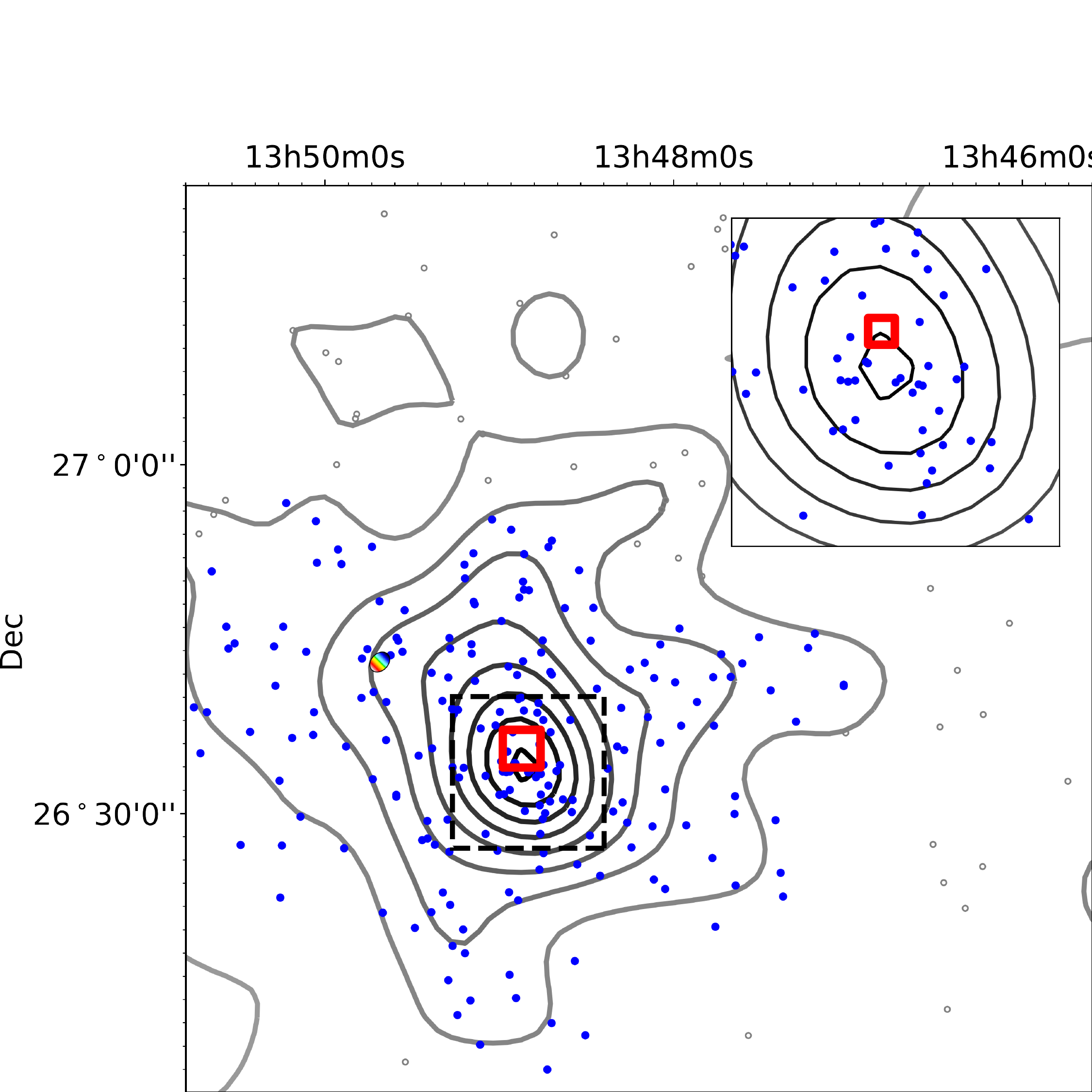}
        \put (20,76) {\fontsize{8}{0}\hl{A1795}}        \put (20,71) {\fontsize{8}{0}\hl{$\mathcal{N}=208$, $z\approx0.067$}}
      \end{overpic}
    \end{subfigure}
    \hspace{-0.7em}
    \begin{subfigure}[b]{0.48\linewidth}
      \begin{overpic}[width=\linewidth]{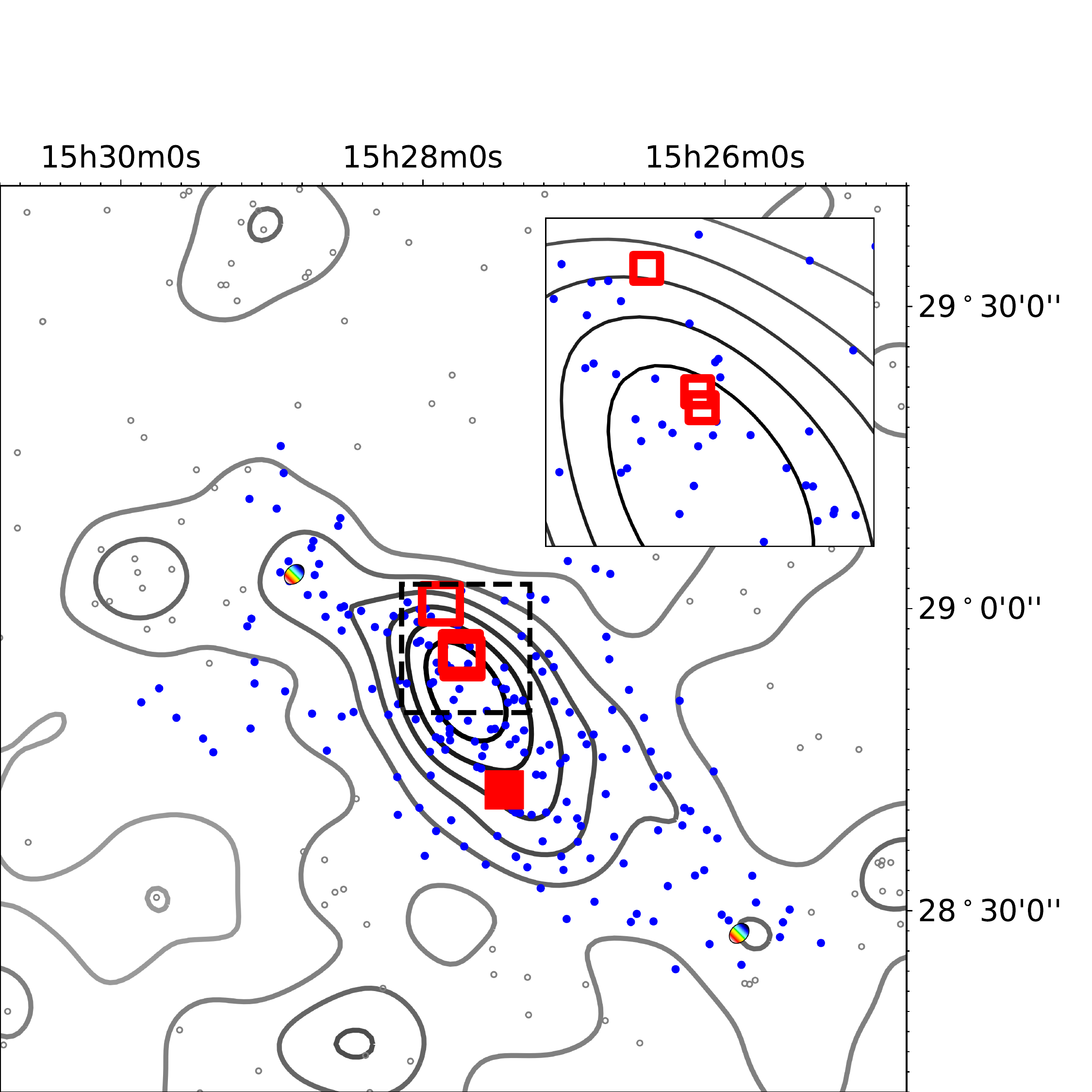}
        \put (3,76) {\fontsize{8}{0}\hl{A2079}}        \put (3,71) {\fontsize{8}{0}\hl{$\mathcal{N}=202$, $z\approx0.07$}}
      \end{overpic}
    \end{subfigure}
    \hspace{-0.7em}
    \begin{subfigure}[b]{0.48\linewidth}
      \begin{overpic}[width=\linewidth]{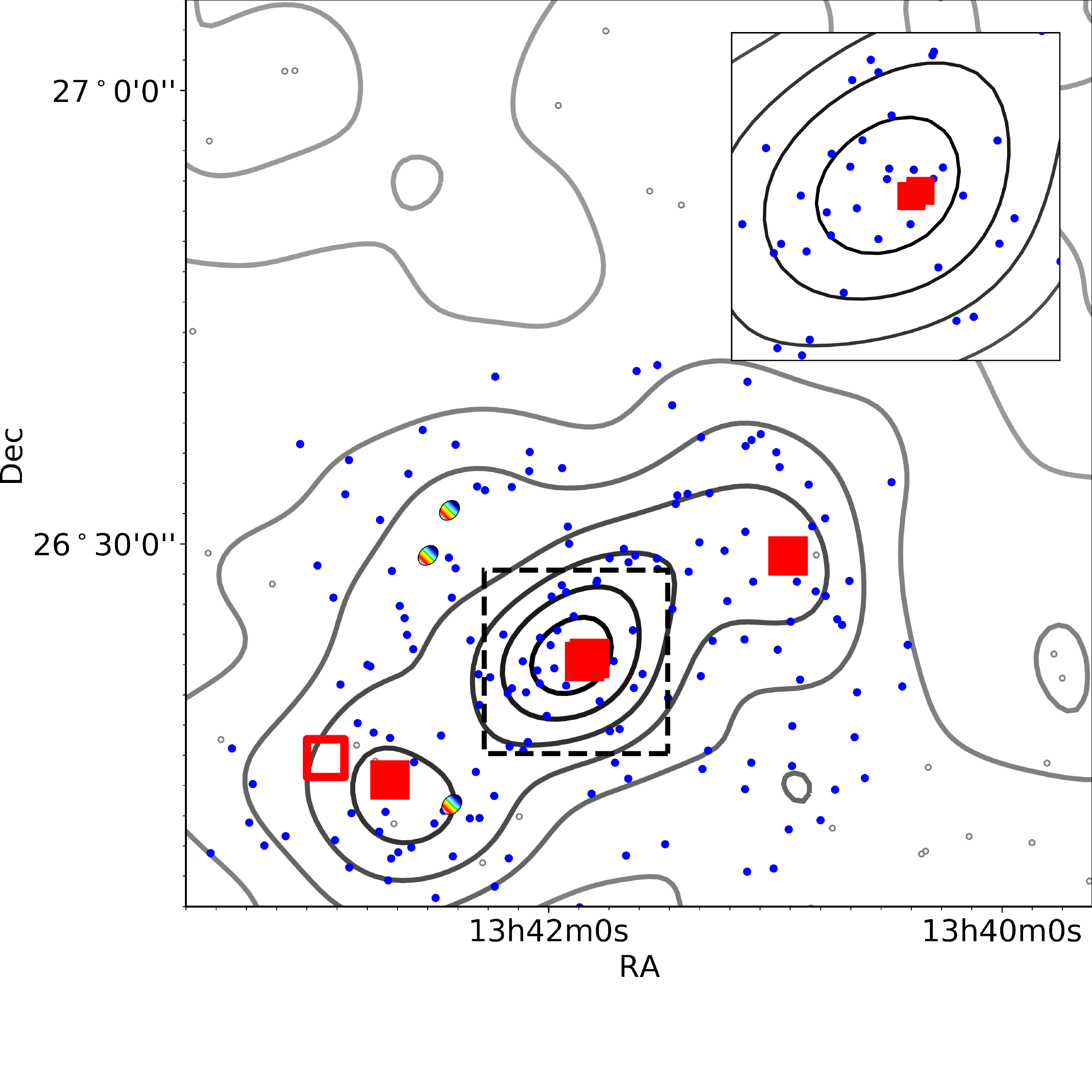}
        \put (20,93) {\fontsize{8}{0}\hl{A1775}}        \put (20,88) {\fontsize{8}{0}\hl{$\mathcal{N}=185$, $z\approx0.075$}}
         \put (25,35) {\fontsize{8}{0}\hl{A}}		\put (45,50) {\fontsize{8}{0}\hl{B}}		\put (63,60) {\fontsize{8}{0}\hl{C}}
      \end{overpic}
    \end{subfigure}
    \hspace{-0.7em}
    \begin{subfigure}[b]{0.48\linewidth}
      \begin{overpic}[width=\linewidth]{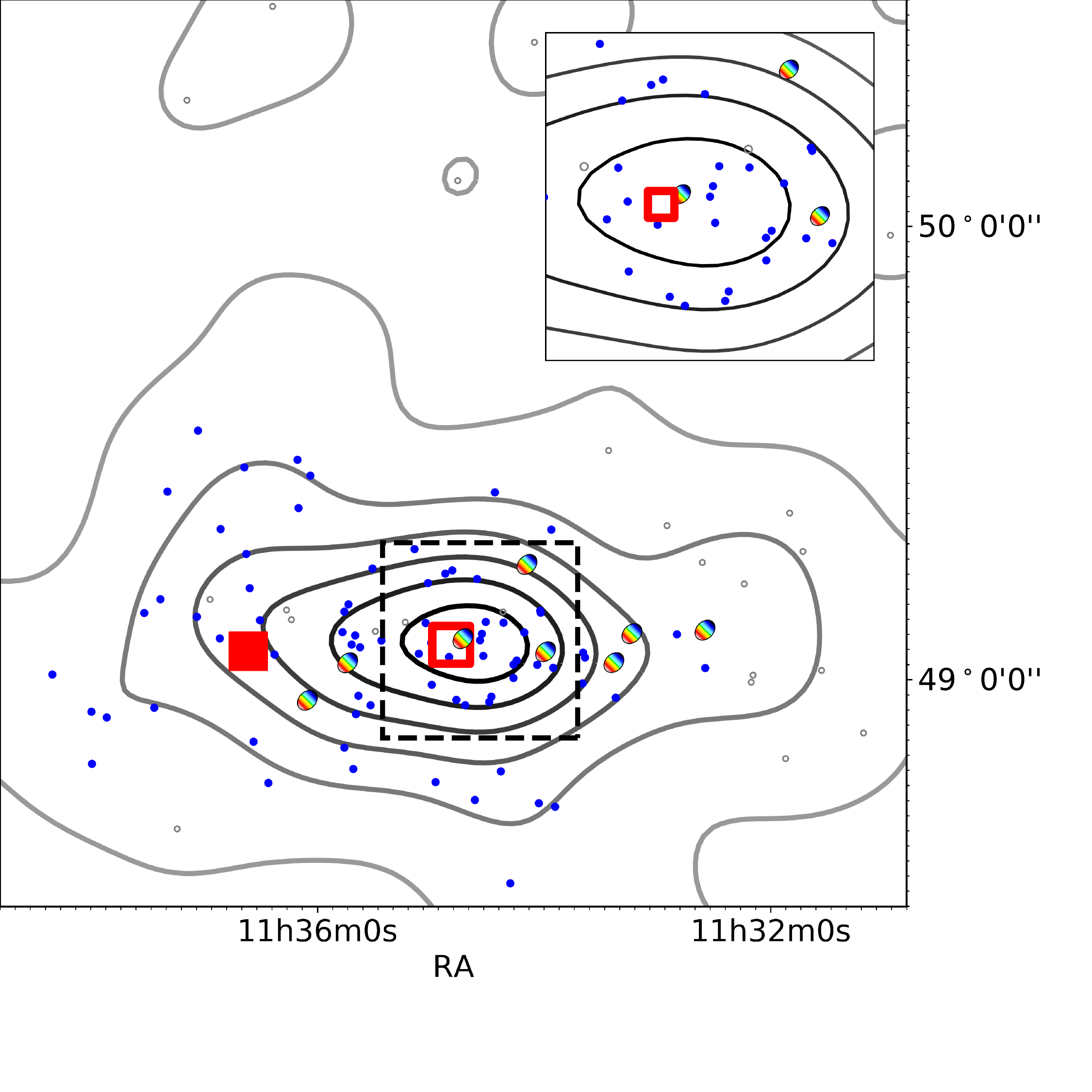}
        \put (3,93) {\fontsize{8}{0}\hl{A1314}}        \put (3,88) {\fontsize{8}{0}\hl{$\mathcal{N}=85$, $z\approx0.032$}}
      \end{overpic}
    \end{subfigure}
    \hspace{-0.7em}
    \vspace{-1em}
\caption[Galaxy clusters where at least one galaxy has been observed with kinematics: A1795, A2079, A1775, A1314]{The same as Figure 1. \textit{Top Left}: A1795 \citep{breugel1984a1795, buote1996clusters, fabian2001a1795, kokotanekov2018a1795}. This cluster lies within the Bo{\"o}tes supercluster. \textit{Top right}: A2079. This cluster is located within the CBS. \textit{Bottom Left}: A1775. This cluster has three components which are labelled \citep{oegerle1995clusters, kopylov2009a1775, zhang2011a1775}. The two BCGs of A1175B appear to be very close but the easternmost of the pair is closer in redshift to A1775A \citep{kopylov2009a1775}. A1775A and A1775B may be interacting as suggested by X-ray observations of the hot gas \citep{andersson2009clusters}. \textit{Bottom Right}: A1314. This low-mass cluster has a disturbed morphology and is likely to be merging \citep{wilbur2018clusters}.}
\end{figure*}

\bibliographystyle{mnras}
\bibliography{MasterBibliography} 

\end{document}